\documentclass[a4paper,12pt, british]{article}

\pdfoutput=1

\usepackage{amsmath}
\usepackage[]{graphicx}

\usepackage[british]{babel}
\usepackage[utf8]{inputenc}
\usepackage{pdflscape}
\usepackage{enumerate}
\usepackage{amsbsy}
\usepackage{amsmath} 
\usepackage[]{graphics}
\usepackage{mathrsfs}
\usepackage{wrapfig}
\usepackage{mathtools}
\usepackage[normalem]{ulem}

\usepackage{amsfonts}
\usepackage{pstricks}
\usepackage{color}
\usepackage{setspace}

\usepackage{jcappub_ver}			
\usepackage[noabbrev]{cleveref}		

\crefformat{equation}{(#2#1#3)}
\crefrangeformat{equation}{(#3#1#4) to~(#5#2#6)}
\crefmultiformat{equation}{(#2#1#3)}{ and~(#2#1#3)}{, (#2#1#3)}{ and~(#2#1#3)}

\newcommand{\bea}{\begin{eqnarray}} \newcommand{\eea}{\end{eqnarray}}
\newcommand{\el}{\nonumber \\}
\newcommand{\re}[1]{(\ref{#1})}

\newcommand{\pat}{\partial}

\newcommand{\para}{\paragraph}

\renewcommand{\a}{\alpha}
\renewcommand{\b}{\beta}

\newcommand{\ha}{\frac{1}{2}}

\newcommand{\rmd}{\mathrm{d}}

\newcommand{\ie}{i.e.\ }

\newcommand{\half}{\ha}

\newcommand{\Mpl}{M_{{}_{\mathrm{Pl}}}}

\newcommand{\bigO}{\mathcal{O}}

\title{Higgs-\texorpdfstring{$R^2$}{R\textasciicircum 2} inflation -- full slow-roll study at tree-level}

\author[a]{Vera-Maria~Enckell,}
\author[a]{Kari~Enqvist,}
\author[a,b]{Syksy~R{\"a}s{\"a}nen}
\author[a]{and Lumi-Pyry Wahlman}

\affiliation[a]{University of Helsinki, Department of Physics and Helsinki Institute of Physics,\\ P.O. Box 64, FIN-00014 University of Helsinki, Finland}

\affiliation[b]{Birzeit University, Department of Physics \\
P.O. Box 14, Birzeit, West Bank, Palestine}

\emailAdd{vera-maria.enckell@helsinki.fi}
\emailAdd{kari.enqvist@helsinki.fi}
\emailAdd{syksy.rasanen@iki.fi}
\emailAdd{pyry.wahlman@helsinki.fi}

\abstract{
We consider Higgs inflation with an $\a R^2$ term. It adds a new scalar degree of freedom, which leads to a two-field model of inflation. We do a complete slow-roll analysis of the three-dimensional parameter space of the $R^2$ coefficient $\a$, the non-minimal coupling $\xi$ and the Higgs self-coupling $\lambda$.
We find three classes of inflationary solutions, but only pure $R^2$ and attractor solutions fit observations.
We find that pure Higgs inflation is impossible when the $R^2$ term is present regardless of how small $\a$ is. However, we can have Higgs-like inflation, where the amplitude of the perturbations does not depend on $\a$ and the predictions as a function of e-folds are the same as in Higgs inflation, although the inflationary trajectory is curved in field space.
The spectral index is $0.939 < n_R < 0.967$, and constraining it to the observed range, the tensor-to-scalar ratio varies from $3.8\times10^{-3}$ to the maximum allowed by observations, $0.079$. Observational constraints on isocurvature perturbations contribute to these limits, whereas non-Gaussianity is automatically in the range allowed by observations.}

\begin{document}

\begin{flushleft}
	\hfill		 HIP-2018-37/TH \\
\end{flushleft}

\maketitle
  
\setcounter{tocdepth}{2}

\setcounter{secnumdepth}{3}

\section{Introduction} \label{sec:intro}

Higgs inflation, based on the only fundamental scalar field in the Standard Model (SM) of particle physics, is an appealingly simple model of inflation. The potential of the SM Higgs without gravity is too steep \cite{Isidori:2007vm, Hamada:2013mya, Fairbairn:2014nxa}, but successful inflation is achieved with the non-minimal coupling $\xi H^{\dagger}H R$ of the Higgs doublet $H$ to the Ricci scalar $R$ \cite{Bezrukov:2007ep}. Such a coupling is generated by quantum corrections \cite{Callan:1970ze}. When this term is taken into account, the amplitude of perturbations is right for $\xi=4\times10^4 \sqrt{\lambda}$ (where $\lambda$ is the Higgs quartic coupling), and the tree-level predictions for the scalar spectral index $n_R=0.96$, its running and the tensor-to-scalar ratio $r=5\times10^{-3}$ agree well with observations \cite{Planck2018}. However, quantum corrections complicate the picture. In most cases, loop corrections do not change the qualitative behaviour \cite{Fumagalli:2016lls}, though for tuned values of the model parameters the Higgs potential may develop an inflection point \cite{Allison:2013uaa, Bezrukov:2014bra, Hamada:2014iga, Hamada:2014wna, Bezrukov:2014ipa, Rubio:2015zia, Fumagalli:2016lls, Enckell:2016xse, Bezrukov:2017dyv, Rasanen:2017, Ezquiaga:2017fvi, Rasanen:2018a}, a hilltop \cite{Fumagalli:2016lls, Rasanen:2017, Enckell:2018a} or a degenerate vacuum \cite{Jinno:2017a, Jinno:2017b}, which can significantly affect predictions.

In curved spacetime quantum corrections, in addition to changing the effective Higgs potential as in flat spacetime, also generate higher order curvature terms. Indeed, the first inflationary model, by Alexei Starobinsky, was based on such terms produced by the trace anomaly \cite{Starobinsky:1980te}. In general, Riemann tensor terms that do not reduce to a function of the Ricci scalar (or higher order Lovelock invariants) suffer from the Ostrogradski instability \cite{Simon:1990ic, MuellerHoissen:1990, Woodard:2006nt}. As is well known, in the metric formulation of general relativity an $F(R)$ action is equivalent to Einstein--Hilbert gravity plus a scalar field at the classical level \cite{Sotiriou:2006}, and also at one loop on-shell \cite{Ohta:2017, Ruf:2017}. If we restrict to terms of up to dimension four, this reasoning leads to (neglecting the cosmological constant) $F(R)=\Mpl^2 R+\a R^2$, which is at the background level equivalent to the original trace anomaly model of Starobinsky. The amplitude of perturbations is right for $\a=8\times10^8$, and the predictions as a function of e-folds are the same as in Higgs inflation (although the number of e-folds is different due to lower reheating temperature \cite{Bezrukov:2011gp}).

When both the non-minimal coupling of the Higgs field and the $R^2$ term are taken into account, we obtain a two-field model of inflation. Indeed, renormalisation group running implies that, absent fine-tuning, $8\pi^2 \a \gtrsim (\xi+\frac{1}{6})^2$ (different sources have a slightly different numerical coefficient) \cite{Salvio:2015kka, Calmet:2016fsr, Salvio:2017oyf, Ghilencea:2018rqg}. Models of inflation with a non-minimally coupled scalar field and the $R^2$ term have been discussed in \cite{Barbon:2015, Salvio:2015kka, Salvio:2017oyf, Kaneda:2015jma, Calmet:2016fsr, Wang:2017fuy, Ema:2017rqn, Pi:2017gih, He:2018gyf, Gorbunov:2018llf, Ghilencea:2018rqg, Wang:2018kly, Gundhi:2018wyz, Karam:2018, Enckell:2018b, Antoniadis:2018ywb, Kubo:2018} (see also \cite{Kofman:1985aw}).

The effect of the $R^2$ term in an extended Higgs inflation model was first discussed in passing in \cite{Barbon:2015}. In \cite{Kaneda:2015jma} an $h$-dependent coupling to the $R^2$ term was considered, focusing on the stability of Starobinsky inflation in the case $\lambda=0$.
In \cite{Wang:2017fuy} the authors found the attractor solutions, studied the fit to data and estimated the isocurvature contribution.
The attractor solutions were also studied in \cite{Ema:2017rqn, He:2018gyf}, and it was found that the spectral index and tensor-to-scalar ratio are close to the pure Higgs or Starobinsky case and isocurvature perturbations are small. In the recent work \cite{Gundhi:2018wyz} the authors studied different inflationary trajectories analytically, including isocurvature effects, and found the merging of different inflationary regions and the possibility of large $r$.
The effect of the $R^2$ term on renormalisation group running was studied in \cite{Salvio:2015kka, Salvio:2017oyf, Calmet:2016fsr}, with emphasis on the feature that a large value of $\xi$ will in general induce a large value of $\a\sim\xi^2/(8\pi^2)$ (see also \cite{Netto:2015cba, Liu:2018hno}). In \cite{Gorbunov:2018llf} its effect on the unitarity problem, and via renormalisation group running to vacuum stability, was considered. In \cite{Ghilencea:2018rqg} two-loop corrections were studied in detail, and they found corrections of the order of a few percent.

We perform a full study of the parameter space of the $R^2$ coefficient $\a$, the non-minimal coupling $\xi$ and the Higgs self-coupling $\lambda$. We first study all slow-roll regions analytically and then scan them numerically, including observational constraints on the amplitude, spectral index, its running, tensor-to-scalar ratio, non-Gaussianity and isocurvature perturbations. We find that pure Higgs inflation (meaning the case where the inflationary trajectory is aligned with the Higgs direction) is not possible, and identify an inflationary region that mixes the two scalar fields and gives a range of predictions in agreement with the current observational data. In the region where the two fields mix we can have Higgs-like inflation, where the amplitude of the perturbations does not depend on $\a$ and the predictions as a function of e-folds are the same as in Higgs inflation.

In \cref{sec:set-up} we introduce the action, rewrite it in terms of two scalar fields and perform a conformal transformation to the Einstein frame. We use the frame-covariant formalism and show that we do not recover pure Higgs inflation. We also review the CMB observables and their observational ranges. In \cref{sec:res} we first look at the slow-roll regions analytically and then present our numerical results and compare to previous work. In \cref{sec:conc} we summarise our conclusions.

\section{The set-up} \label{sec:set-up}

\subsection{Action and changes of variables} \label{sec:action}

We consider an action with a non-minimally coupled scalar field, which we identify with the SM Higgs, and a gravitational sector with an $F(R)$ term. We reparametrise the $F(R)$ term to write the extra degree of freedom in terms of a scalar field, redefine fields and make a conformal transformation to the Einstein frame.

We begin from the following action for scalar field $h$ and metric $g_{\a\b}$:
\bea \label{action1}
  S &=& \int\rmd^4 x \sqrt{-g} \left[ \ha F(R) + \ha G(h) R - \ha g^{\a\b} \pat_\a h \pat_\b h - V(h) \right] \ ,
\eea
where $g$ is the determinant of $g_{\a\b}$ and $R$ is the Ricci scalar, determined by the metric and its first two derivatives. The usual Higgs inflation corresponds to \cite{Bezrukov:2007ep}
\bea \label{Higgs-inflation}
  \text{Higgs inflation}:\quad F(R)=\Mpl R,\quad G(h)=\xi h^2,\quad V(h)=\frac{\lambda}{4} {(h^2-v^2)}^2.
\eea
In order to exchange $F(R)$ for $R$ plus a scalar field, we rewrite the action \re{action1} as \cite{Sotiriou:2008}
\begin{align} \label{action2}
  S &= \int\rmd^4 x \sqrt{-g} \left[ \ha G(h) R + \ha \{ F(\phi) + F'(\phi) (R - \phi) \} - \ha g^{\a\b} \pat_\a h \pat_\b h - V(h) \right] \ ,
\end{align}
where $\phi$ is an auxiliary field. Variation with respect to $\phi$ yields (if $F''\neq 0$) $\phi=R$, and if we substitute this result back to \eqref{action2} we recover the original action \eqref{action1}. By redefining the auxiliary field as $\varphi \equiv F'(\phi)$ we obtain the action
\bea \label{action3}
  S &=& \int\rmd^4 x \sqrt{-g} \left[ \ha [ \varphi + G(h) ] R - W(\varphi) - \ha g^{\a\b} \pat_\a h \pat_\b h - V(h) \right] \ ,
\eea
where the potential for $\varphi$ is
\bea \label{W}
  W(\varphi) \equiv \ha \{ \phi(\varphi) \varphi - F[\phi(\varphi)] \} \ .
\eea

To obtain minimally coupled fields, we make a conformal transformation to Einstein frame
\bea
  g_{\a\b} \rightarrow \Omega^2 g_{\a\b} = [ \varphi + G(h) ] g_{\a\b} \ .
\eea
The action then becomes
\bea \label{action4}
  S &=& \int\rmd^4 x \sqrt{-g} \left[ \ha R - \frac{3}{4 \Omega^4} g^{\a\b} ( \pat_\a \varphi \pat_\b \varphi + 2 \pat_\a \varphi \pat_\b G + \pat_\a G \pat_\b G ) \right. \el
  && - \left. \ha \frac{1}{\Omega^2} g^{\a\b} \pat_\a h \pat_\b h - \hat V(h, \varphi) \right] \ ,
\eea
where we have chosen units such that the Planck mass is unity, and the new potential reads
\bea \label{Vhat1}
  \hat V(h, \varphi) = \frac{V(h) + W(\varphi)}{\Omega^4} \ .
\eea
Note that $\Omega^2>0$ is a necessary and sufficient condition for neither of the scalar fields to be ghosts.

Restricting to dimension four terms, we have
\bea
F(R)=R+\a R^2,\quad G(h)=\xi h^2,\quad V(h)=\frac{\lambda}{4} (h^2-v^2)^2,
\eea
where $\alpha$ and $\xi$ are constants. For this choice the action becomes
\bea \label{action5}
  S &=& \int\rmd^4 x \sqrt{-g} \left[ \ha R - \frac{3}{4 \Omega^4} g^{\a\b} \pat_\a \varphi \pat_\b \varphi - \frac{3\xi h}{\Omega^4} g^{\a\b} \pat_\a \varphi \pat_\b h \right. \el
  && \left. - \frac{\Omega^2 + 6 \xi^2 h^2}{2 \Omega^4} g^{\a\b} \pat_\a h \pat_\b h - \hat V(h, \varphi) \right] \ ,
\eea
and the potential is
\bea \label{Vhat2}
  \hat V(h, \varphi) = \frac{\lambda}{4} \frac{(h^2 - v^2)^2}{\Omega^4} + \frac{1}{8 \a} \frac{(\varphi-1)^2}{\Omega^4} \ .
\eea
If $\a$ is negative, the electroweak (EW) vacuum at $h=v, \varphi=1$ is a saddle point, so we take $\a>0$. The kinetic cross-term could be removed with the field redefinition $\Phi = \Omega^2 = \varphi + \xi h^2$. However, we cannot have a canonical kinetic term for both fields at the same time, hence we have to anyway take into account curvature in field space, which we do with the frame-covariant formalism.

\subsection{Frame-covariant formalism and the Higgs inflation limit} \label{sec:fcf}

In multi-field inflation, all fields cannot be simultaneously canonically normalised if the field space is curved. This is the case in Higgs--Starobinsky inflation, and various choices of field coordinates have been used in the literature \cite{Barbon:2015, Kaneda:2015jma, Wang:2017fuy, Ema:2017rqn, Pi:2017gih, He:2018gyf, Gorbunov:2018llf, Ghilencea:2018rqg, Wang:2018kly, Gundhi:2018wyz}. 
The exact physical quantities are independent of the chosen frame, but we have to be careful when making the slow-roll approximation. We follow the frame-covariant formalism introduced in \cite{Karamitsos:2017elm, Karamitsos:2018lur} where the slow-roll parameters are frame-covariant, ensuring that quantities expanded in them are invariant under frame and field reparametrisation, \ie independent of the choice coordinates in the space of the metric and the scalar fields.

As noted above, we use the Einstein frame, which simplifies many of the equations. The field space metric of the action \re{action5} reads
\begin{align}\label{field metric}
  G_{AB} = \frac{1}{\Omega^4} \left( \begin{array}{cc}
    3/2 & 3 \xi h \\
    3 \xi h & \Omega^2 + 6 \xi^2 h^2
  \end{array} \right) \ ,
\end{align}
where the indices $A,B$ refer to field coordinates that are collectively denoted as
\bea
  \phi^A=(\varphi,h) \ .
\eea
The inverse of $G_{AB}$ is $G^{AB}$, and the two tensors are used to lower and raise indices in field space.

The first two slow-roll parameters can be expressed as\footnote{This definition of $\eta$ differs by a sign from that in \cite{Karamitsos:2017elm, Karamitsos:2018lur}, as we find the latter inconsistent with the equations of the observables $r$, $n_R$, and $\a_R$. Using our convention, in the single field case $\epsilon$ and $\eta$ relate to the usual slow-roll parameters as $\epsilon=\epsilon_V, \eta=2\eta_V-4\epsilon_V$, with $\epsilon_V=\ha (V'/V)^2, \eta_V=V''/V$. This sign issue propagates to the higher order slow-roll parameters $\epsilon_n$.}
\begin{align}
  \epsilon & = \half \frac{G^{AB} V_{,A} V_{,B}}{V^2} \notag \\
  \eta & = G^{AB} \frac{V_{,A}}{V} \frac{\epsilon_{,B}}{\epsilon} \ , \label{eq:sr-def}
\end{align}
and in general
\begin{align}\label{epsilon n+1}
  \epsilon_{n+1} = G^{AB} \frac{V_{,A}}{V} \frac{(\epsilon_n)_{,B}}{\epsilon_n} \ .
\end{align}
The general expressions for $\epsilon$ and $\eta$ are complicated. In the limit $h\ll1, v\ll1$ we get
\begin{align}
  \epsilon & = \frac{4}{3 (\varphi - 1)^2} + \bigO(h^2) \\
  \eta & = - \frac{8 \varphi}{3 (\varphi - 1)^2} + \bigO(h^2) \ ,
\end{align}
which is the usual result for Starobinsky inflation. However, in the limit $h\gg1$ we have (for any value of $\varphi$)
\begin{align}
  \epsilon & = \frac{4}{3} (1+6\xi) + \bigO\left( \frac{1}{h^2} \right) \\
  \eta & = \frac{4 \xi}{3 \lambda \a} [ - 6 \lambda \a + \xi (1 + 6 \xi) ] + \bigO\left( \frac{1}{h^2} \right) \ ,
\end{align}
so the slow-roll parameters do not reduce to those of Higgs inflation even if $\varphi=1$. Furthermore a constant $\varphi$ is never a solution to the field equations, which implies that pure Higgs inflation becomes unstable when we add the $R^2$ term. (In contrast, it is consistent to put $h=0$ in the region where the $R^2$ term dominates if we neglect $v$, whose effect is insignificant in the inflationary region.) We do not recover pure Higgs inflation in the limit $\a\to0$ either (unless $\a$ is exactly zero, in which case there is only one scalar degree of freedom, the Higgs field), as we will show in \cref{sec:ana}.

\subsection{Inflationary predictions and CMB observables} \label{sec:obs}

\para{Predictions for adiabatic and tensor perturbations.}

Expanded to lowest order in slow-roll parameters \eqref{eq:sr-def}, the scalar spectral index, the running of spectral index and the tensor-to-scalar ratio read
\bea \label{rnsalfa}
  n_R & = & 1 -2\epsilon + \eta - \sin (2\Theta)\mathcal{D}_N T_{\mathcal{R}\mathcal{S}}\\
  \a_R & = & - 2 \epsilon\eta -\eta\zeta - 2 \cos(2\Theta)\cos^2 \Theta {(\mathcal{D}_N T_{\mathcal{R}\mathcal{S}})}^2 + \sin (2\Theta) \mathcal{D}_N \mathcal{D}_N T_{\mathcal{R}\mathcal{S}} \\
    r & = & 16 \epsilon \cos^2\Theta \ ,
\eea
where $\mathcal{D}_N$ is the frame covariant derivative with respect to the number of e-folds (see (2.29) in \cite{Karamitsos:2017elm}), $T_{\mathcal{R}\mathcal{S}}$ is the transfer function between the curvature and isocurvature modes and $\Theta$ is the transfer angle. The last two are discussed in more detail below.
Non-Gaussianity, estimated in terms of the amplitude of the bispectrum defined with the three-point correlator of the comoving curvature perturbation, is
\begin{equation}\label{fNL}
f_{NL}=\frac{5}{6}\frac{N^{,A}N^{,B} \nabla_A\nabla_B N}{{(N_{,A}N^{,A})}^2} \ ,
\end{equation}
where $\nabla_A$ is the covariant derivative in field space.

\para{Predictions for isocurvature perturbations.}

Let us then consider isocurvature perturbations. The formal solutions for the comoving curvature and isocurvature perturbations are given in terms of the transfer functions $T_{\mathcal{R}\mathcal{S}}$, $T_{\mathcal{S}\mathcal{S}}$ as
\bea
  \mathcal{R}(t)&=&\mathcal{R}(t_*) + T_{\mathcal{R}\mathcal{S}}(t_*,t) \mathcal{S} (t_*)\\
  \mathcal{S}(t)&=&\mathcal{R}(t_*) + T_{\mathcal{S}\mathcal{S}}(t_*,t) \mathcal{S} (t_*) \ ,
\eea
where star denotes quantities calculated at the pivot scale $k_*=0.05$ ${\rm Mpc}^{-1}$. The solution for the transfer functions in terms of the number of e-folds $N$ reads
\begin{eqnarray}\label{Trs}
  T_{\mathcal{R}\mathcal{S}}(N_*,N) &=& -\int_{N_*}^N dN' A(N') T_{\mathcal{S}\mathcal{S}}(N_*,N')\\
  T_{\mathcal{S}\mathcal{S}}(N_*,N) &=& \exp \left[-\int_{N_*}^N dN' B(N')\right] \ ,
\end{eqnarray}
where the functions $A(N)$, $B(N)$ depend on the model of inflation, and can be written in terms of various frame-covariant parameters \cite{Karamitsos:2017elm, Karamitsos:2018lur}. The effect of the energy transfer between the curvature and isocurvature modes can then be encoded in the transfer angle $\Theta$\footnote{The transfer angle $\Theta$ used in \cite{Karamitsos:2017elm, Karamitsos:2018lur} is related to the correlation angle $\Delta$ used in the Planck analysis \cite{Ade:2015lrj} as $\Theta=\Delta+\pi/2$.}
\bea\label{theta}
  \cos\Theta = \frac{1}{\sqrt{1+T_{\mathcal{R}\mathcal{S}}^2}} 
\eea
and the isocurvature fraction
\bea\label{isoc}
  \beta_{\rm iso}=\frac{T_{\mathcal{S}\mathcal{S}}^2}{1+T_{\mathcal{R}\mathcal{S}}^2+T_{\mathcal{S}\mathcal{S}}^2} \ .
\eea

\para{Observational values.}

When comparing to observations we apply the Planck ranges for the cosmological observables at the pivot scale $k_*=0.05$ ${\rm Mpc}^{-1}$. The Hubble exit of this scale corresponds to (assuming that reheating lasts 4 e-folds, as in pure Higgs inflation \cite{Figueroa:2009jw,Figueroa:2015,Repond:2016}, although see \cite{Ema:2016, DeCross:2016, Sfakianakis:2018lzf})
\begin{equation} \label{efolds}
  N_* = 52 - \frac{1}{4} \ln \frac{r_*}{0.079} \ ,
\end{equation}
which is also the constraint for the minimum amount of inflation. The number of e-folds depends on the tensor-to-scalar ratio $r_*$ at the pivot scale for which we use the CMB constraint \cite{Planck2018}
\begin{equation}\label{obs:r}
  r_* < 0.079 \ .
\end{equation}
For the amplitude of scalar perturbations we use the mean value \cite{Planck2018}
\begin{equation} \label{obs:amplitude}
  24 \pi^2 A_*=\frac{V_*}{\epsilon_*\cos^2 \Theta_*}=4.97 \times 10^{-7} \ .
\end{equation}
The 95\% confidence limits for the spectral index read \cite{Planck2018}
\begin{equation} \label{obs:ns}
  0.9554 < n_R < 0.9726 \ ,
\end{equation}
and the running of the spectral index is constrained to
\begin{equation} \label{obs:as}
  - 0.0207 < \a_R < 0.0065 \ .  
\end{equation}
The limits for the transfer angle $\Theta$ are given by \cite{Planck2018}
\begin{equation} \label{obs:theta}
 -0.25 < \sin \Theta < 0.23 \ ,
\end{equation}
and the upper bound of the isocurvature fraction is
\begin{equation} \label{obs:beta-iso}
 \beta_{\rm iso} < 0.38 \ .
\end{equation}
For some choices of datasets and likelihoods, the Planck data in fact gives a lower bound on $ \beta_{\rm iso}$ \cite{Planck2018}. If confirmed, this would rule out most of our allowed region on the $(n_R, r)$ plane. Finally, non-Gaussianity is constrained as \cite{Ade:2015ava}
\begin{equation} \label{obs:fnl}
  - 9.2 < f_{NL} < 10.8 \ .
\end{equation}

\section{Slow-roll regions} \label{sec:res}

\subsection{Analytical treatment} \label{sec:ana}

We find that the viable slow-roll area in field space divides into two parts. The first one is the Starobinsky region where $h\ll 1$ and $\varphi\gg 1$. The second one is a region centered around a positively curved parabola in field space. The value of $\xi$ determines whether these regions are attractors or repellers. Let us look at the regions analytically before going to the numerical results.

Let us first consider the case when $\xi$ is positive. There are then four slow-roll regions, two of which can lead to successful inflation. The first region is the one around the parabola
\begin{align}
  \varphi = 1 + b + c h^2 \ , \label{eq:par-c}
\end{align}
with
\begin{align}
  b & = - \frac{2 \lambda \a}{12 \lambda \a \xi + \xi^2 (1 + 6 \xi)} = \frac{c}{(c+d)(1-6d)} \el
  c & = \frac{2 \lambda \a}{\xi} \el
  d & = \xi + \frac{1}{6} \ .
\end{align}
This region is an attractor: the field rolls towards the parabola and then along it to the EW minimum. The parameter $c$ determines the location of this region while $d$ controls its width.

\begin{figure}[t]
  \center
  \begin{minipage}[t]{0.45\textwidth}
    \center
    \includegraphics[width=\textwidth]{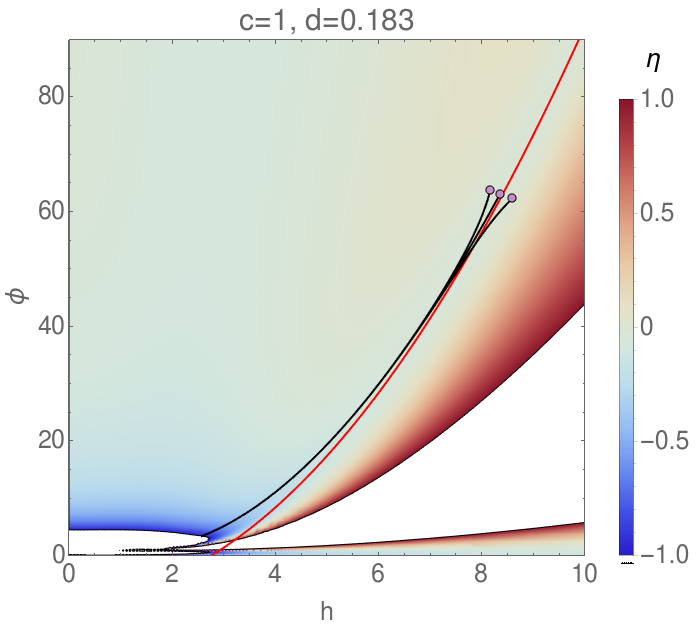}
    \caption{Field trajectories for \mbox{$c=1$} and \mbox{$d=0.18$} (\mbox{$\xi=2\lambda\a\approx0.01$}). We show only trajectories for which the observables are in the allowed ranges \cref{obs:ns,obs:as,obs:theta,obs:beta-iso,obs:fnl}. Only solutions near the attractor satisfy these criteria. The red line is the analytical approximation \re{eq:par-c} for the attractor, which is not very good for the small value of $\xi$ used. The coloured region shows the area where \mbox{$|\epsilon|, |\eta|<1$}.}
    \label{fig:pos-xi-s}
  \end{minipage}
  \hfill
  \begin{minipage}[t]{0.45\textwidth}
    \center
    \includegraphics[width=\textwidth]{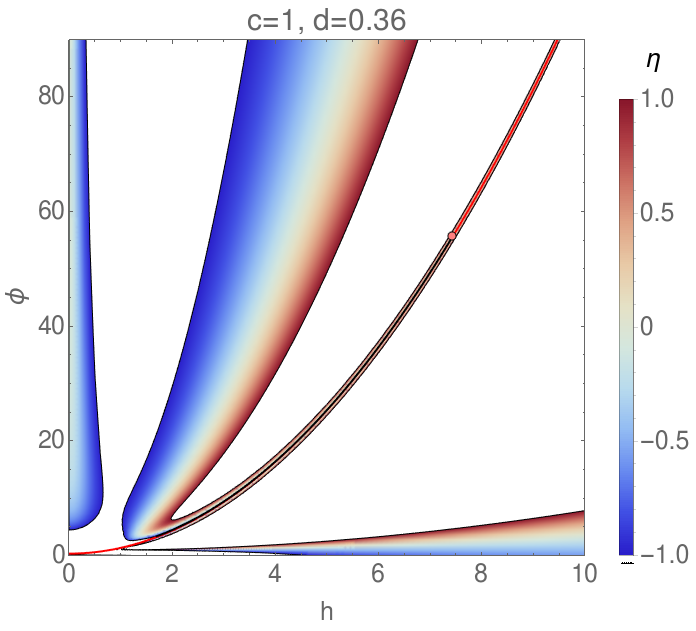}
    \caption{Field trajectories for \mbox{$c=1$} and \mbox{$d=0.36$} (\mbox{$\xi=2\lambda\a\approx0.2$}). We show only trajectories for which the observables are in the allowed ranges \cref{obs:ns,obs:as,obs:theta,obs:beta-iso,obs:fnl}. The red line is the analytical approximation \re{eq:par-c} for the attractor. We have chosen a small value of $\xi$ for representation purposes (for larger $\xi$ the region around the attractor becomes too thin). The coloured region shows the area where \mbox{$|\epsilon|, |\eta|<1$}.}
    \label{fig:pos-xi-m}
  \end{minipage}
\end{figure}

The second region is the Starobinsky region. It corresponds to a hill in the $h$ direction and is hence a repeller. The field rolls away from the line \mbox{$h=0$} towards the attractor \eqref{eq:par-c} and then along it to the EW minimum. For $\xi \gtrsim 1$, the slow-roll region around the Starobinsky solution is narrow and yields typically less than one e-fold of inflation. Hence it does not give a viable model (except in the fine-tuned case when we are right on the Starobinsky solution). However, for $\xi \ll 1$ the Starobinsky region grows wider and merges with the region around the parabola. We then have a broad connected slow-roll region, giving a third class of inflationary solutions in addition to the attractor and Starobinsky inflation. There are successful initial field values also near the line $h=0$. However, the observational constraints on the transfer angle \cref{obs:theta} restrict the initial field values either to a region close to the attractor
or essentially to the Starobinsky solution at $h=0$. This situation is illustrated in \cref{fig:pos-xi-s}, with field trajectories starting from different points in field space. We show in \cref{fig:pos-xi-m} how the viable region becomes smaller and disconnected with increasing $\xi$.

For positive $\xi$, the remaining non-successful slow-roll regions correspond to a region near the parabola
\begin{align}
  \varphi = - d h^2 \ . \label{eq:par-d}
\end{align}
and a region where $h \ll1$ and $\varphi < 0$. Note that in these regions $\Omega^2<0$, so the scalar fields are ghosts. The first region is neither an attractor nor a repeller while the second is a repeller. In both cases the field rolls away from the EW minimum.

\begin{wrapfigure}{R}{0.5\textwidth}
  \center
  \vspace{-10pt}
  \begin{minipage}{0.45\textwidth}
    \center
    \includegraphics[scale=0.3]{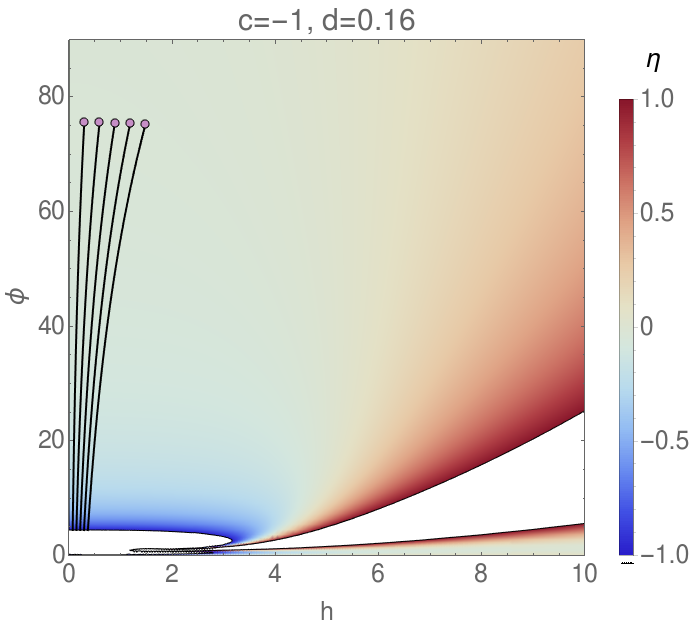}
    \caption{Field trajectories for \mbox{$c=-1$} and \mbox{$d=0.16$} (\mbox{$\xi=2\lambda\a\approx-7 \times 10^{-3}$}) in the slow-roll region. We show only trajectories for which the observables are in the allowed ranges \cref{obs:ns,obs:as,obs:theta,obs:beta-iso,obs:fnl}. The coloured region shows the area where \mbox{$|\epsilon|, |\eta|<1$}.}
    \label{fig:neg-xi}
  \end{minipage}
\end{wrapfigure}

Let us then consider the case when $\xi$ is negative. Now the Starobinsky region is the only successful slow-roll region. The Starobinsky solution is an attractor and the field rolls towards it, as shown in \cref{fig:neg-xi}. The region around the parabola \eqref{eq:par-c}, which now opens downward, is a repeller. The second parabola \eqref{eq:par-d} has a negative slope for values above the conformal value $-1/6 < \xi < 0$ and a positive slope for more negative values. In both cases the region around it is neither an attractor nor a repeller. Finally, the region with $h \ll 1$ and $\varphi < 0$ is an attractor, but the field rolls away from the EW minimum.

\para{Higgs inflation limit.}

Let us now take a closer look at solutions on the attractor \eqref{eq:par-c}. When $c$ decreases, the parabola becomes wider. In the $c \to 0$ limit it becomes the line $\varphi=1$, which we might expect to correspond to pure Higgs inflation, given that the action \re{action5} reduces to the usual Higgs case in this limit. When the field lies on the parabola, we can calculate the slow-roll parameters analytically. In the limit $d h^2 \gg 1$ (\ie $\xi h^2 \gg 1$), when $c>1$ we obtain
\bea
  \epsilon & = & \frac{9(c+d)}{2 (6 c + 6 d - 1) N^2} + \bigO(N^{-3}) \\
  \eta & = & - \frac{2}{N} + \bigO(N^{-2}) \ .
\eea
It might seem that the predictions reduce to the pure Higgs inflation case for $d\gg c$.
For $c \gtrsim 1$ this is indeed the case, but inflation is not driven by the Higgs field alone (the field rolls also in the $\varphi$ direction).
We do not recover pure Higgs inflation in the limit $c\to0$ (\ie $\alpha\to0$) either. For $c\ll1$ the slow-roll parameters are, to leading order in $c$,
\bea
  \epsilon & = & \frac{9(c+d)}{2 (6 c + 6 d - 1) N^2} + \bigO(c) \\
  \eta & = & \frac{3 d}{2 c (1-6d)^2 N^2} + \bigO(1) \label{eq:eta-pole} \ .
\eea
We see that $\eta$ diverges as $c$ approaches zero.\footnote{In \cite{Wang:2017fuy} it was claimed that inflation along the valley approaches the pure Higgs case when $\Mpl/M \to 0$ ($c \to 0$ in our notation). However, when the $c \to 0$ limit is taken before the $\xi h^2 \gg 1$ limit (in pure Higgs inflation $\xi h^2$ is finite but $c=0$), the divergence \eqref{eq:eta-pole} persists.}
The divergence is due to the fact that the second derivative of the potential with respect to $\varphi$ is divergent. This derivative is present in all of the higher order slow-roll parameters, which diverge as $1/c$. As a result the running of the spectral index diverges as $1/c^2$, and higher order corrections diverge even faster. This implies that pure Higgs inflation is impossible when the $R^2$ term is present in the action regardless of the value of $\a$.\footnote{One could argue that the analytical approximation breaks down in the limit $\xi\to0$. However, our numerical results show that while the attractor approaches the line $\varphi=1$ in this limit, the slow-roll parameters become larger than one.} Because the theory with $\a\neq0$ has an extra scalar degree of freedom, it is not a small perturbation of the case $\a=0$, regardless of how small $\a$ is. The limit $\a\to0$ is singular, and it does not commute with solving the equations of motion, as is also the case in pure Starobinsky inflation \cite{Simon:1990ic}.

However, as long as $1\lesssim c \ll d$, we can still have Higgs-like inflation at the attractor. We define a solution to be Higgs-like if the amplitude of perturbations does not depend on $\a$ and the predictions as a function of e-folds are the same as in pure Higgs inflation. However, the trajectory is a parabola in field space, instead of a line with constant $\varphi$. Similarly, we define a solution to be Starobinsky-like if the amplitude is determined by $\a$ and the predictions as a function of e-folds are the same as in pure Starobinsky inflation.

\subsection{Numerical results} \label{sec:num}

We have performed a comprehensive scan of the parameter space of the slow-roll regions. Overall we have six parameters ($\a$, $\xi$, $\lambda$, $N$ and two field coordinates at the pivot scale) and three constraints for the field evolution (the constraint on e-folds, the amplitude at the pivot scale and the condition that at the end of inflation either $|\epsilon|=1$ or $|\eta|=1$, whichever occurs first\footnote{These do not correspond to the usual conditions for the end of slow-roll in single-field inflation, but give a slightly later point in time. The difference in the number of e-folds depends on the parameters, varying from negligibly small to $1.4$, comparable to the effect of reheating.}). When written in terms of the parameters $c$ and $d$, the slow-roll parameters and the number of e-folds are independent of $\lambda$ and the amplitude is linear in $\lambda$. Hence we keep $c$, $d$ and one of the field coordinates at the pivot scale as free parameters, and solve $\lambda$ from the normalisation condition. For the numerical analysis we chose some representative values of $c$ and varied the field coordinate and the parameter $d$. We checked that the results converge to analytical estimates in various limits to make sure that we have included all relevant values in our scan.

\begin{figure}[t]
  \center
  \begin{minipage}[t]{0.45\textwidth}
    \center
    \includegraphics[width=\textwidth]{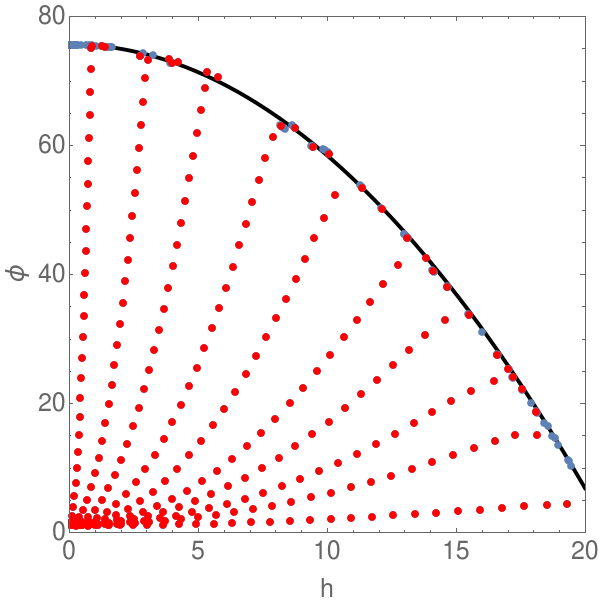}
    \caption{Numerical results in field space. Each dot corresponds to the pivot scale for one solution that satisfies the observational constraints on $r$, $n_R$, $\a_R$, $\Theta$ and $f_{NL}$. Red points denote solutions that lie on the attractor \eqref{eq:par-c}. The black curve is a negatively curved parabola given by \mbox{$\varphi = 76 - 0.17 h^2$}, and it is approximately the $\xi \to 0$ limit. The EW minimum is in the bottom left corner. For representation purposes we have included only some of the datapoints used in \cref{fig:rns}.}
    \label{fig:field-space}
  \end{minipage}
  \hfill
  \begin{minipage}[t]{0.45\textwidth}
    \center
    \includegraphics[width=\textwidth]{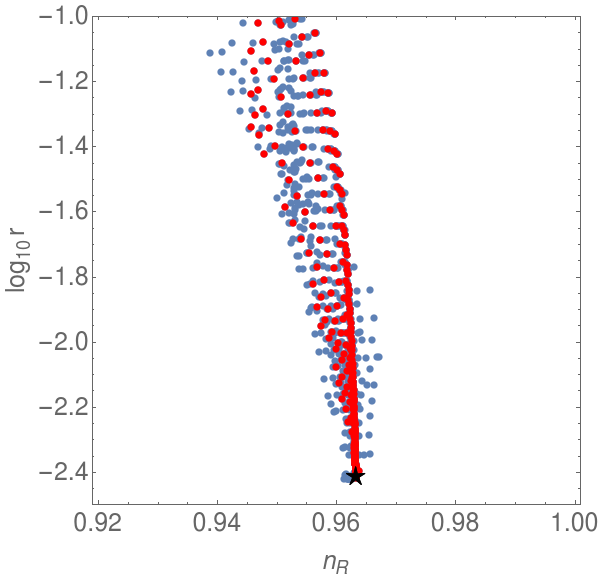}
    \caption{Numerical results for spectral index $n_R$ and tensor-to-scalar ratio $r$. Red points denote solutions that lie on the attractor \eqref{eq:par-c}. The black star is the pure Starobinsky solution, whose predictions agree with pure Higgs inflation. All values of $r$ from the Starobinsky case \mbox{$r = 3.8 \times 10^{-3}$} up to the observational upper limit \mbox{$r < 0.079$} are allowed, and the spectral index is \mbox{$0.939 < n_R < 0.967$}. Other observables are not single-valued on this plane.}
    \label{fig:rns}
  \end{minipage}
\end{figure}

Our results are illustrated in \cref{fig:field-space,fig:rns}. \Cref{fig:field-space} shows the solutions in field space. Each dot corresponds to the pivot scale for one solution that satisfies the above three constraints and also agrees with all of the observational constraints \cref{obs:as,obs:theta,obs:beta-iso,obs:fnl}. The solutions fall into three categories: those near the Starobinsky solution (top of the black curve at $h=0$), those on the downward opening parabola (traced by the black curve), and those on the paths going down from the black curve towards the EW minimum on the bottom left corner.

The first category corresponds to Starobinsky inflation. When $|\xi| \ll 1$, the Starobinsky region becomes wider (for $\xi=0$ the first correction to the slow-roll parameters is $\bigO(\lambda \a h^4)$ as opposed to $\bigO(\xi h^2)$), and we can have Starobinsky-like solutions even for $h \gg 1$. However, only solutions with $h \lesssim 1$ are allowed by the constraint on the transfer angle \eqref{obs:theta}.

The second category corresponds to solutions near the attractor with a small $\xi$. When the value of $\xi$ becomes small enough, the region around the attractor merges with the region around the Starobinsky solution, and the field can roll from the Starobinsky region to the attractor, while maintaining slow-roll. However, the constraint on the transfer angle restricts the field to values that are basically at the attractor. For these solutions, the tensor-to-scalar ratio can be large, $r \sim 0.1$, and they constitute most of the allowed region in \cref{fig:rns}.

The third category corresponds to solutions near the attractor when $\xi \gtrsim 0.1$, with different paths having different values of $c=2\lambda\a/\xi$ (recall that $c$ determines the location of the attractor). The values of $d$ and $\lambda$ increase when approaching the EW minimum on the bottom left corner. The spectral index is the same as in the pure Starobinsky case, but $r$ can be larger for $c \ll 1$ and $d \lesssim 1$ (note that the analytical approximation \eqref{eq:par-c} for the attractor breaks down at small values of $c$ and the exact attractor does not approach the line $\varphi=1$).

\Cref{fig:rns} shows our results for the spectral index $n_R$ and the tensor-to-scalar ratio $r$. Each dot represents again a single solution, and red points denote solutions on the attractor \eqref{eq:par-c}. We only show the allowed region, because the parameters and observables are degenerate (i.e. we can obtain the same predictions for $r$ and $n_R$ with multiple parameter values). We have constrained $\a_s$, $f_{NL}$ and $\sin \Theta$ to agree with the observed values. The black star shows the tree-level prediction of pure Higgs inflation (which coincides with the prediction of pure Starobinsky inflation). The solutions around the attractor in the case of small $\xi$ constitute a large part of the allowed region. The spectral index $n_R$ is in the range $0.939 < n_R < 0.967$ and  tensor-to-scalar ratio $r$ ranges from the pure Starobinsky value $3.8\times10^{-3}$ to the current observational upper limit $r < 0.079$, considering all inflationary regions. The attractor solution has $0.946 < n_R < 0.964$ and the same range of $r$. The values of $r$ close to the upper limit are obtained near the attractor when $\xi \ll 1$. These numbers can be compared to the analytical predictions of pure tree-level Higgs or Starobinsky inflation: $n_R=0.96$ and $r=5\times10^{-3}$.

Non-Gaussianity is small, $-9.4 \times 10^{-5}<f_{NL}<0.021$. For positive $\xi$, the isocurvature fraction is $\beta_{\rm iso}<0.017$, while for negative $\xi$, $\beta_{\rm iso}$ can reach $0.24$. Tightening the observational constraint on $\Theta$ would narrow down the allowed region in the $n_R$ direction, but $r$ could still be large. Note that if all forms of matter produced at preheating reach thermal equilibrium, isocurvature perturbations are converted to adiabatic perturbations and these limits disappear.

\subsection{Comparison to previous work} \label{sec:com}

Let us compare our results to previous papers that have studied non-minimally coupled scalar field inflation with the $R^2$ term, concentrating on those that use the SM Higgs potential. In \cite{Wang:2017fuy} the authors found the attractor solutions and studied the fit to the data, finding that it is possible to obtain $r>0.03$. They also estimated the isocurvature contribution, finding it can be large when $\xi$ and (in our notation) $c$ are small, which agrees with our results. In \cite{He:2018gyf}, the attractor solution was also studied, finding that the results for the spectral index and tensor-to-scalar ratio are basically the same as in the pure Higgs or Starobinsky case and isocurvature perturbations are small. Our results for the attractor solution agree with \cite{Wang:2017fuy, Ema:2017rqn, He:2018gyf}, when all the constraints are taken into account.

In the recent work \cite{Gundhi:2018wyz}, which appeared as we were preparing this paper for publication, the authors studied different inflationary trajectories analytically, including the isocurvature contribution, backed by numerical studies in some cases. They paid particular attention to the effect of initial conditions, finding that the field quickly settles onto the slow-roll trajectory, though with possible interesting wiggles when inflation starts from a hilltop. They also noted that the limit $\a\to0$ is singular and found the merging of the two inflationary regions in the limit of small $\xi$, and the possibility of large $r$. We have done a comprehensive numerical study of the slow-roll trajectories, and compared to observational constraints in detail, including the running of the spectral index and non-Gaussianity. Where our work overlaps, we are in agreement.

\section{Conclusions} \label{sec:conc}

We have done a systematic study of inflation in the case when both the non-minimally coupled Standard Model Higgs field and the $R^2$ term are present. We have analysed all slow-roll regions, examined the full parameter space and taken into account constraints from observations.

There are three classes of solutions: Starobinsky inflation driven by the $R^2$ term, an attractor region that is curved in field space, and a third region where these two merge together at small values of the non-minimal coupling $\xi$. The solutions in the third class are excluded by the constraints on the transfer angle \eqref{obs:theta}. We find that Higgs--Starobinsky inflation can produce a spectral index bounded as $0.939 < n_R < 0.967$ (on the attractor $0.946 < n_R < 0.964$) and, constraining $n_R$ to the observed range, the tensor-to-scalar ratio varies from $r=3.8\times 10^{-3}$ up to the current upper limit $r < 0.079$, reaching the upper limit when $\xi \ll 1$. These solutions also agree with the observational limits on the amplitude of perturbations, running of the spectral index, isocurvature fraction, isocurvature transfer angle and non-Gaussianity. Of these, the isocurvature fraction and non-Gaussianity are automatically within the observational range and provide no extra constraints, whereas the running of the spectral index and isocurvature transfer angle contribute to the limits. The results extend the analytical prediction of pure tree-level Higgs or Starobinsky inflation, $n_R=0.96$, $r=5\times10^{-3}$. We also find that $\xi$ can be much smaller than in pure Higgs inflation (all points shown in \cref{fig:rns} can be reached with $\xi<1$), which can help with the unitarity issue of Higgs inflation \cite{Barbon:2009ya, Burgess:2009ea, Burgess:2010zq, Lerner:2009na, Lerner:2010mq, Hertzberg:2010dc, Bauer:2010, Bezrukov:2010jz, Bezrukov:2011sz, Calmet:2013hia, Weenink:2010rr, Lerner:2011it, Prokopec:2012ug, Xianyu:2013, Prokopec:2014iya, Ren:2014, Escriva:2016cwl, Fumagalli:2017cdo, Gorbunov:2018llf}.

Extending previous studies, we have taken into account the observational limits for the running of the spectral index, non-Gaussianity and isocurvature in detail. Notably, we find that there is no pure Higgs inflation limit when the $R^2$ term is present, no matter how small its coefficient $\a$ is. This is related to the fact that the limit $\a\to0$ is singular, as noted in \cite{He:2018gyf, Gundhi:2018wyz}. One way to avoid this destabilisation is to use the Palatini formulation of general relativity, where the $R^2$ term does not generate a new scalar degree of freedom \cite{Sotiriou:2006, Enckell:2018b, Antoniadis:2018ywb}.

We have relied on the standard description of reheating for the pure Higgs inflation case \cite{Figueroa:2009jw,Figueroa:2015,Repond:2016}. The reheating temperature is lower in Starobinsky inflation \cite{Bezrukov:2011gp}, reducing the number of e-folds. Reheating in the mixed Higgs--Starobinsky model requires a dedicated study, continuing along the lines of \cite{He:2018mgb}. We expect changes in reheating to affect the allowed region in the model parameter space (i.e. the values of $\a$, $\lambda$ and $\xi$ may change), but the overall picture remains the same: parameters can be found to match the observations. The main difference would likely be a small shift in the allowed region: for longer reheating $n_R$ is smaller and $r$ is larger.

A more careful analysis would also take into account quantum corrections. In \cite{Ghilencea:2018rqg} small corrections at the level of a few percent, rising with $\lambda$, were found at two loops. Standard Model particles other than the radial Higgs mode (the top quark in particular) should also be included in the analysis, because they affect the running of the Higgs self-coupling, as in pure Higgs inflation.

\acknowledgments

We thank Tommi Tenkanen who participated in the beginning of this project. LPW is supported by the Magnus Ehrnrooth foundation.

\bibliographystyle{JHEP}
\bibliography{starmet}


\end{document}